\documentclass[10pt]{article}

\usepackage{graphicx}
\usepackage{graphics}
\usepackage{amscd}
\usepackage{amsmath}
\usepackage{amsfonts}
\usepackage{amssymb}
\usepackage{enumerate}

\usepackage[none]{hyphenat}

\usepackage{hyperref}

\font\daude=cmcsc10 at 18pt
\font\hoato=cmcsc10 at 12pt
\font\hoa=cmcsc10 at 12pt

\newcommand{\F}{\mathbb{F}}
\newcommand{\Z}{\mathbb{Z}}

\newcommand{\xn}{x^n - 1}

\newcommand{\pf}{{\bf Proof : }}
\newtheorem{theorem}{Theorem}

\newtheorem{definition}[theorem]{Definition}

\begin{document}

\begin{center}
{\daude {Construction of cyclic DNA codes over the Ring
\vskip 7pt		
\huge{$\Z_4[u]/\langle u^2-1 \rangle $}} Based on the deletion distance}
\end{center}

\vskip 20pt
\begin{center} {{\hoa Sukhamoy Pattanayak\footnote{Department of Applied Mathematics, Indian School of Mines, Dhanbad, India. {\tt sukhamoy88@gmail.com}} and 
Abhay Kumar Singh\footnote{Department of Applied Mathematics, Indian School of Mines, Dhanbad, India. {\tt singh.ak.am@ismdhanbad.ac.in}} 
}}
\end{center} 
\vskip 6pt

\vskip 10pt
\leftskip 0.9in
\rightskip 0.9in
\noindent 
{\bf Abstract.} In this paper, we develop the theory for constructing DNA cyclic codes of odd length over $R=\Z_4[u]/\langle u^2-1 \rangle$ based on the deletion distance. Firstly, we relate DNA pairs with a special 16 elements of ring $R$. Cyclic codes of odd length over $R$ satisfy the reverse constraint and the reverse-complement constraint are discussed in this paper. We also study the $GC$-content of these codes and their deletion distance. The paper concludes with some examples of cyclic DNA codes with $GC$-content and their respective deletion distance.


\leftskip 0.0in
\rightskip 0.0in

\vskip 30pt

\centerline {{\hoato 1. Introduction}}

\vskip 10pt
Deoxyribonucleic acid (DNA) contains genetic instructions for the structure and biological developments of life. It has the information on how the biological cell runs, reproduces builds and repair itself. DNA strands sequences consists of four nucleotides; two purines: adenine ($A$) and guanine ($G$), and two pyrimidines: thymine ($T$) and cytosine ($C$). The two strands of DNA are linked with a rule that are name as Watson-Crick complement (WCC). According to WCC rule; every (A) is linked with a (T), and every (C) with a (G), and vice versa. We write this is as $\overline{A}=T, \overline{T}=A, \overline{G}=C$ and $\overline{C}=G$. For example, if $x=(GCATAG)$, then its complement is $\overline{x}=(CGTATC)$.

\vskip 6pt
DNA computing links genetic data analysis with scientific computations in order to handle computationally difficult problems. Leonard Adleman \cite{ad} introduced an experiment involving the use of DNA molecules to solve a hard computational problem in a test tube. His study was based on the WCC property of DNA strands. Several paper have been discussed different techniques to construct a set of DNA codewords that are unlikely to form undesirable bonds with each other by hybridization. For example, in \cite{mar}, four different constraints i.e., \emph{the Hamming constraint}, the \emph{reverse constraint}, the \emph{reverse-complement constraint}, the \emph{fixed $GC$-content constraint} on DNA codes are considered. The first three constraints is to avoid undesirable hybridization between different strands. The fixed $GC$-content which ensures all codewords have similar thermodynamic characteristic.

\vskip 6pt
The design of DNA strands has several applications in genetics and bioengineering. For example, application of biomolecular computing is the design of DNA chips for mutational analysis and for sequence. DNA strands are designed so that each strand uniquely hybridizes with its WCC sequence and not to any other sequence. DNA strands focuses on constructing large sets of DNA codewords with prescribed minimum Hamming distance. Recently D'yachkov et al. discussed the deletion similarity distance which is more suitable then the Hamming distance in \cite{Dy, Dy1}.

\vskip 6pt
Furthermore, cyclic DNA computing has generated great interest because of their more storage capacity than silicon based computing systems, and this motivates many authors to study it. Since then, the construction of DNA cyclic codes have been discussed by several Authors in \cite{ab, ba, be}, \cite{ga}-\cite{lia}, \cite{si, pa, yi2}. Gaborit and King discussed linear construction of DNA codes in \cite{ga}. In \cite{ab}, DNA codes over finite field with four elements were studied by Abualrub et al. Later, Siap et al. considered DNA codes over the finite ring $\F_2[u]/\langle u^2-1 \rangle$ with four element in \cite{si}. In \cite{lia}, Liang and Wang discussed cyclic DNA codes over four element ring $\F_2+u\F_2$. Yildiz and Siap in \cite{yi2} studied the ring with 16 elements over $\F_2[u]/\langle u^4-1 \rangle$ and their structure for DNA computing. In \cite{ba}, codes over the ring $F_4+vF_4, v^2=v$ with 16 elements are considered by Bayram et al. and discussed some DNA application. Zhu and Chen studied cyclic DNA codes over the non-chain ring $\F_2[u,v]/\langle u^2,v^2-v,uv-vu \rangle$ in \cite{zh}. In \cite{be}, Bennenni et al. considered the chain ring $\F_2[u]/\langle u^6 \rangle$ with 64 elements and discussed DNA cyclic codes over this ring.

\vskip 6pt
Codes over the ring $\Z_4+u\Z_4,~u^2=0$ have been discussed in the papers \cite{yi, yi1}. Later, Ozena et al. in \cite{oze} consider cyclic and some constacyclic codes over the ring $\Z_4[u]/\langle u^2-1 \rangle$. Here, we study a construction of DNA cyclic codes of a finite non-chain ring $R$ with 16 elements.

\vskip 6pt
The rest of the paper is organized as follows. The next section present the structure of the ring $\Z_4+u\Z_4,~u^2=1$ as well as some basic description and definition of cyclic DNA codes. Also we establish a 1-1 correspondence $\theta$ between the elements of the ring $R$ and DNA double pair. We study cyclic codes satisfy the reverse constraint over $R$ in section 3. In section 4, we also discuss cyclic codes satisfy the reverse-complement constraint over $R$. We study the $GC$-content and the deletion distance of DNA cyclic codes over $R$ in section 5. In section 6, we discuss the examples of cyclic DNA cyclic codes of different length with deletion distance and $GC$-content. Section 7 concludes the paper.

\vskip 30pt
\centerline {{\hoato 2. Preliminaries}}

\vskip 6pt
Let $R$ be a commutative, characteristic 4 ring $\Z_4+u\Z_4=\{a+ub\vert a,b \in \Z_4 \}$ with $u^2=1$. $R$ can also be thought of as the quotient ring $ \Z_4[u]/\langle u^2-1 \rangle$. The unit elements of $R$ are $\{ a+ub :\text{either}~ a~ \text{or}~ b~ \text{is a unit in}~ \Z_4\}$.
R has a total of seven ideals given by
\begin{center}
	$\{0\},\langle 2u \rangle,\langle 1+u \rangle,\langle 3+u \rangle,\langle 2+2u \rangle,\langle 2u,1+u \rangle ~\text{and}~R$.
\end{center}
$R$ is a non-principal local ring with the maximum ideal $\langle 2u,1+u \rangle$. Here, $R$ is a non-chain ring. 

\vskip 6pt
A cyclic code $C$ of length $n$ over $R$ is a $R$-submodule of $R^n$ which is invariant with respect to shift operator that maps a codeword $c=(c_0, c_1, \ldots , c_{n-1}) \in C$ to another codeword $(c_{n-1}, c_0, \ldots , c_{n-2})$ in $C$. It is well known that $C$ is cyclic code of length $n$ over $R$ if and only if $C$ is an ideal in the ring $R_n=R[x]/\langle x^n-1\rangle$. The Hamming weight of a codeword $c=(c_0,c_1,\ldots,c_{n-1})$ denoted by $w_H(c)$ is the number of non zero entries in $c$. The Hamming distance $d(c_1, c_2)$ between two codewords $c_1$ and $c_2$ is the Hamming weight of the codeword $c_1-c_2$. 

\vskip 6pt
Let $S_{D_4}=\{A,T,G,C\}$ represent the DNA alphabet. We define a DNA code of length $n$ to be a set of codewords $(x_0,x_1,\ldots, x_{n-1})$, where $x_i \in \{A,T,G,C\}$. These codewords must satisfy the four constraints mentioned in the introduction. We use the same notation for the set
\begin{center}
	$S_{D_{16}}=\{AA,TT,GG,CC,AT,TA,GC,CG,GT,TG,AC,CA,CT,TC,AG,GA\},$
\end{center}
which is originally presented in \cite{ozt}. 
Since the commutative ring $R$ is of the cardinality 16, then we define the map $\theta$ which gives a one-to-one correspondence between the elements of $R$ and DNA double pairs $S_{D_{16}}$, which are given in Table 1. The elements $\{0,1,2,3\}$ of $\Z_4$ are in one-to-one correspondence with the nucleotide DNA bases $S_{D_4}=\{A,T,G,C\}$ such that
$0\rightarrow A, 1 \rightarrow T, 2 \rightarrow G ~\text{and}~ 3 \rightarrow C$. The codons satisfy the Watson-Crick complement which is given by $\overline{A}=T, \overline{T}=A, \overline{G}=C, \overline{C}=G$. \\
Let $x=x_0x_1 \ldots x_{n-1} \in R^n$ be a vector. The reverse of $x$ is defined as $x^r = x_{n-1}x_{n-2} \ldots x_1x_0$, the complement of $x$
is $x^c =\overline{x_0}~\overline{x_1}\ldots \overline{x_{n-1}} $, and the reverse-complement, also called the Watson-Crick complement (WCC) is defined as $x^{rc} =\overline{x_{n-1}}~\overline{x_{n-2}} \ldots \overline{x_1}~\overline{x_0} $.
\begin{definition}
	A linear code $C$ of length $n$ over $R$ is said to be \bfseries{reversible} if $x^r \in C~~ \forall~ x \in C$, \bfseries{complement} if $x^c \in C~~ \forall~ x \in C$ and \bfseries{reversible-complement} if $x^{rc} \in C~~ \forall~ x \in C$.	
\end{definition}
\begin{definition}[D'yachkov et al. \cite{Dy,Dy1}]
	Let $1\leq D \leq n-1$ be a fixed integer. A cyclic code $C$ of length $n$ is called an $(n,D)$ DNA code of distance $D$ over $R$ if
	\begin{enumerate}[{\rm (i)}]
		\item $C$ is a cyclic code, i.e., $C$ is an ideal of $R[x]/\langle x^n-1\rangle$ $;$
		\item For any codeword $X \in C, X \neq X^{rc}$	and $X^{rc} \in C$ $;$
		\item $S(X,Y)\leq n-D-1$, for any $X,Y \in C,~X \neq Y$.
	\end{enumerate}
\end{definition}
\begin{definition}
	Let $C$ be a code over $R$ of arbitrary length $n$ and $c \in C$ be a codeword where $c=(c_0,c_1,\cdots,c_{n-1}), c_i \in R$, then we define 
	$$\Phi(c):C\longrightarrow S_{D_4}^{2n},$$ $(a_0+ub_0,a_1+ub_1,\ldots,a_{n-1}+ub_{n-1})\longrightarrow (a_0,a_1,\ldots,a_{n-1},b_0,b_1,\ldots,b_{n-1})$.
\end{definition}

\begin{center}
	{\bf Table 1.} $\theta$-table for DNA correspondence \\~\\
	\begin{tabular}{c c c}
		\hline
		Elements 
		of $R$ & Gray image & DNA Codons \\
		\hline
		$0$ & $(0,0)$ & $AA$ \\
		$u$ & $(0,1)$ & $AT$  \\
		$2u$ & $(0,2)$ & $AG$ \\
		$3u$ & $(0,3)$ & $AC$   \\
		
		$1$ & $(1,0)$ & $TA$  \\
		$1+u$ & $(1,1)$ & $TT$   \\
		$1+2u$ & $(1,2)$ & $TG$  \\
		$1+3u$ & $(1,3)$ & $TC$   \\
		
		$2$ & $(2,0)$ & $GA$  \\
		$2+u$ & $(2,1)$ & $GT$  \\
		$2+2u$ & $(2,2)$ & $GG$  \\
		$2+3u$ & $(2,3)$ & $GC$    \\
		
		$3$ & $(3,0)$ & $CA$  \\
		$3+u$ & $(3,1)$ & $CT$  \\
		$3+2u$ & $(3,2)$ & $CG$ \\
		$3+3u$ & $(3,3)$ & $CC$ \\
		\hline
		
	\end{tabular}
\end{center}

\vskip 6pt

For each polynomial $f(x)=a_0+a_1x+\cdots+a_rx^r$ with $a_r\neq 0$, we define the reciprocal of $f(x)$ to be the polynomial
\begin{center}
	$f^*(x)=x^rf(1/x)=a_r+a_{r-1}x+\cdots+a_0x^r$.
\end{center}
It is easy to see that $~deg~(f ^*(x)) \leq deg~(f (x))$ and if $a_0 \neq 0$, then $~deg~(f ^*(x)) = deg~(f (x))$. $f(x)$ is called a self-reciprocal polynomial if there is a constant $m$ such that $f^*(x) = mf(x)$.

\vskip 6pt
The structure of a cyclic code of odd length $n$ over $R$ has been briefly studied in \cite{oze} by the projection map $\varphi (a+ub)=a-b$ from $R$ to $\Z_4$, which is given as following theorem.

\vskip 6pt
\noindent {\bf Theorem 2.1.} {\it Let $n$ be an odd integer and $C$ be a cyclic code of length $n$ over $R$. Then $C=\langle g_1(x)+(1+u)g_2(x),(1+u)g_3(x) \rangle$, where $g_1(x)$ and $g_3(x)$ are generator polynomials of cyclic codes over $\Z_4$ with $g_3(x)\vert g_1(x)\vert (\xn) ~\text{mod~2}~$ and $g_2(x)$ is a polynomial over $R$. If $g_1(x)=g_3(x)$, then $C=\langle g_1(x)+(1+u)g_2(x) \rangle $.
}

\vskip 30pt
\centerline {{\hoato 3. Reversible Codes over $R$}}

\vskip 6pt
In this section, we study the reverse constraint on cyclic codes of odd length over $R$. First, we give some useful lemma's which use the following theorems.

\vskip 6pt
 \noindent {\bf Lemma 3.1.}~\textnormal{\cite{ab2}}
 	{\it Let $f(x),g(x)$ be any two polynomials in $R$ with $~deg~f(x) \geq ~deg~g(x) $. Then
 	\begin{enumerate}[{\rm (i)}]
 		\item $[f(x)g(x)]^*=f^*(x)g^*(x)$;
 		\item $[f(x)+g(x)]^*=f^*(x)+x^{~deg~f-deg~g}g^*(x)$.
 	\end{enumerate}
 }
 
 \vskip 6pt
 \noindent {\bf Lemma 3.2.}~\textnormal{\cite{mas}}
 	{\it Let $C = \langle g(x)\rangle$ be a cyclic code over $\Z_4$, then $C$ is reversible if and
 	only if $g(x)$ is self-reciprocal.
 }
 
 \vskip 6pt
 \noindent {\bf Theorem 3.3.}
 {\it Let $C=\langle g_1(x)+(1+u)g_2(x) \rangle$ be a cyclic code of odd length $n$ over $R$. Then $C$ is reversible if and only if
 	\begin{enumerate}[{\rm (i)}]
 		\item $g_1(x)$ is self-reciprocal;
 		\item \begin{enumerate}[{\rm (a)}]
 			\item $x^ig_2^*(x)=g_2(x)$. Or,
 			\item $g_1(x)=x^ig_2^*(x)+g_2(x),
 			~\text{where}~ i=~deg~g_1(x)-deg~g_2(x)$.
 		\end{enumerate}
 	\end{enumerate}
}

\vskip 6pt
 \pf Suppose $C=\langle g_1(x)+(1+u)g_2(x)  \rangle$ is reversible over $R$, then $C$ is reversible over $\Z_4$, then from Lemma 3.2., $g_1(x)$ is self-reciprocal. That implies 
 \begin{align}
 (g_1(x)+(1+u)g_2(x))^* & = g_1^*(x)+(1+u)x^ig_2^*(x) \nonumber \\ 
 & = g_1(x)+(1+u)x^ig_2^*(x) \\
 & = (g_1(x)+(1+u)g_2(x))a(x) \in C,
 \end{align}
 where $i=~deg~g_1(x)-deg~g_2(x)$. Comparing the degree of (1) and (2), we get $a(x)=c$, where $c\in R$. Then
 \begin{equation}
 g_1(x)+(1+u)x^ig_2^*(x)=c.g_1(x)+(1+u)c.g_2(x).
 \end{equation}
 Multiplying $2(1+u)$ both side of (3) we get, $2(1+u)g_1(x)=c.2(1+u)g_1(x)$. Then the value of $c$ may accepted as  $1,3,u,3u,1+2u,2+u,2+3u,3+2u$. If $c=1$, then $x^ig_2^*(x)=g_2(x)$.
 For $c=3$, we get from (3)
 \begin{equation}
 \begin{split}
 & g_1(x)+(1+u)x^ig_2^*(x)=3g_1(x)+3(1+u)g_2(x) \\ \nonumber
 & \Rightarrow 2g_1(x)=3(1+u)x^ig_2^*(x)+3(1+u)g_2(x).
 \end{split}
 \end{equation}
 Multiplying $(1+u)$ both side of the above equation we have, $g_1(x)=x^ig_2^*(x)+g_2(x)$ as $u^2=1$, where $i=~deg~g_1(x)-deg~g_2(x)$. If $c=u$, then $x^ig_2^*(x)=g_2(x)$.
 For $c=3u$, then by (3) we have, 
 \begin{equation}
 \begin{split}
 & g_1(x)+(1+u)x^ig_2^*(x)=3ug_1(x)+3u(1+u)g_2(x) \\ \nonumber
 & \Rightarrow (1+u)g_1(x)=3(1+u)x^ig_2^*(x)+3(1+u)g_2(x).
 \end{split}
 \end{equation} 
 Multiplying $(1+u)$ both side we get, $g_1(x)=x^ig_2^*(x)+g_2(x)$ as $u^2=1$, where $i=~deg~g_1(x)-deg~g_2(x)$.
 If $c=1+2u$, then by (3) we get, 
 \begin{equation}
 \begin{split}
 & g_1(x)+(1+u)x^ig_2^*(x)=(1+2u)g_1(x)+(1+2u)(1+u)g_2(x) \\ \nonumber
 & \Rightarrow 2ug_1(x)=3(1+u)x^ig_2^*(x)+3(1+u)g_2(x).
 \end{split}
 \end{equation}
 Similarly multiplying $(1+u)$ both side of the above equation we have, $g_1(x)=x^ig_2^*(x)+g_2(x)$ as $u^2=1$, where $i=~deg~g_1(x)-deg~g_2(x)$.
 For $c=2+u$, then by (3) we have,
 \begin{equation}
 \begin{split}
 & g_1(x)+(1+u)x^ig_2^*(x)=(2+u)g_1(x)+(2+u)(1+u)g_2(x) \\ \nonumber
 & \Rightarrow 3(1+u)g_1(x)=3(1+u)x^ig_2^*(x)+3(1+u)g_2(x).
 \end{split}
 \end{equation} 
 Therefore, $g_1(x)=x^ig_2^*(x)+g_2(x)$, where $i=~deg~g_1(x)-deg~g_2(x)$.
 If $c=2+3u$, then by (3) we get, 
 \begin{equation}
 \begin{split}
 & g_1(x)+(1+u)x^ig_2^*(x)=(2+3u)g_1(x)+(2+3u)(1+u)g_2(x) \\ \nonumber
 & \Rightarrow (1+3u)g_1(x)+(1+u)x^ig_2^*(x)=(1+u)g_2(x).
 \end{split}
 \end{equation} 
 Multiplying $(1+u)$ both side we get, $x^ig_2^*(x)=g_2(x)$, where $i=~deg~g_1(x)-deg~g_2(x)$. For $c=3+2u$, then by (3) we have,
 \begin{equation}
 \begin{split}
 & g_1(x)+(1+u)x^ig_2^*(x)=(3+2u)g_1(x)+(3+2u)(1+u)g_2(x) \\ \nonumber
 & \Rightarrow 2(1+u)g_1(x)=(1+u)x^ig_2^*(x)+(1+u)g_2(x).
 \end{split}
 \end{equation}
 Multiplying $(1+u)$ both side we have, $x^ig_2^*(x)=g_2(x)$, where $i=~deg~g_1(x)-deg~g_2(x)$.
 On the other hand, we have
 \begin{align}
 (g_1(x)+(1+u)g_2(x))^* & = g_1^*(x)+(1+u)x^ig_2^*(x) \nonumber \\ 
 & = g_1(x)+(1+u)x^ig_2^*(x) \nonumber \\
 & = (g_1(x)+(1+u)g_2(x))c \in C, \nonumber
 \end{align}
 where $c=1,3,u,3u,1+2u,2+u,2+3u ~\text{or}~3+2u$, $i=deg~f_1(x)-deg~f_2(x)$. Hence, $C$ is reversible.
 
 \vskip 6pt
 \noindent {\bf Theorem 3.4.}
 	{\it Let $C=\langle g_1(x)+(1+u)g_2(x),(1+u)g_3(x) \rangle$
 	with $g_3(x)\vert g_1(x)\vert \xn$ and $g_2(x) \in R[x]$ be a cyclic code of odd length $n$ over $R$. Then $C$ is reversible if and only if
 	\begin{enumerate}[{\rm (i)}]
 		\item $g_1(x)$ and $g_3(x)$ are self-reciprocal;
 		\item $g_3(x)\vert (x^ig_2^*(x)-g_2(x))$, where $i=~deg~g_1(x)-deg~g_2(x).$
 	\end{enumerate}
}

\vskip 6pt
 \pf Suppose $C=\langle g_1(x)+(1+u)g_2(x),(1+u)g_3(x) \rangle$ is reversible. Then $\langle g_1(x)\rangle$ and  $\langle g_3(x)\rangle$ are reversible over $\Z_4$, from Lemma 3.2., $\langle g_1(x)\rangle$ and  $\langle g_3(x)\rangle$ are self-reciprocal. Since $C$ is reversible this implies
 \begin{align}
 (g_1(x)+(1+u)g_2(x))^* & = g_1^*(x)+(1+u)x^ig_2^*(x) ~~~~(i=~deg~f_1(x)-deg~f_2(x))\nonumber \\ 
 & = g_1(x)+(1+u)x^ig_2^*(x)\\
 & = (g_1(x)+(1+u)g_2(x))p(x)+(1+u)g_3(x)q(x).
 \end{align}
 Comparing the degree of (4) and (5), we have $p(x)=c$, where $c\in R$. Then
 \begin{equation}
 g_1(x)+(1+u)x^ig_2^*(x)=c.g_1(x)+c.(1+u)g_2(x)+(1+u)g_3(x)q(x).
 \end{equation}
 Hence, $2(1+u)g_1(x)=c.2(1+u)g_1(x)$. Then the value of $c$ may accepted as $c=1,3,u,3u,1+2u,2+u,2+3u,3+2u$.\\ If $c=1$, then by (6) we get, 
 \begin{equation}
 \begin{split}
 & g_1(x)+(1+u)x^ig_2^*(x)=g_1(x)+(1+u)g_2(x)+(1+u)g_3(x)q(x) \\ \nonumber
 & \Rightarrow x^ig_2^*(x)-g_2(x)=g_3(x)q(x).
 \end{split}
 \end{equation}
 Which means $(x^ig_2^*(x)-g_2(x)) \in (g_3(x))$. Therefore, $g_3(x)\vert (x^ig_2^*(x)-g_2(x))$.\\
 For $c=3$, then by (6) we have, 
 \begin{equation}
 \begin{split}
 & g_1(x)+(1+u)x^ig_2^*(x)=3g_1(x)+3(1+u)g_2(x)+(1+u)g_3(x)q(x) \\ \nonumber
 & \Rightarrow 2g_1(x)+(1+u)x^ig_2^*(x)=3(1+u)g_2(x)+(1+u)g_3(x)q(x)\\
 & \Rightarrow x^ig_2^*(x)-g_2(x)=g_3(x)q(x)~~(\text{multiplying} ~(1+u) ~\text{both side}).
 \end{split}
 \end{equation} 
 Which implies $(x^ig_2^*(x)-g_2(x)) \in (g_3(x))$ as $u^2=1$, where $i=~deg~g_1(x)-deg~g_2(x)$.\\ Hence, $g_3(x)\vert (x^ig_2^*(x)-g_2(x))$.
 Similarly for $c=u$ or $c=3+2u$ we get the same result.\\
 If $c=3u$, then by (6) we get, 
 \begin{equation}
 \begin{split}
 & g_1(x)+(1+u)x^ig_2^*(x)=3ug_1(x)+3u(1+u)g_2(x)+(1+u)g_3(x)q(x) \\ \nonumber
 & \Rightarrow (1+u)g_1(x)+(1+u)x^ig_2^*(x)=3(1+u)g_2(x)+(1+u)g_3(x)q(x) \\
 & \Rightarrow 2(1+u)g_1(x)+2(1+u)x^ig_2^*(x)=2(1+u)g_2(x)+2(1+u)g_3(x)q(x)\\
 & \Rightarrow g_1(x)+x^ig_2^*(x)-g_2(x)=g_3(x)q(x). 
 \end{split}
 \end{equation}
 Hence, $g_1(x)+x^ig_2^*(x)-g_2(x) \in (g_3(x))$ as $u^2=1$, where $i=~deg~g_1(x)-deg~g_2(x)$. Therefore, $g_3(x)\vert (g_1(x)+x^ig_2^*(x)-g_2(x))$. Since $g_3(x) \vert g_1(x)$, we have $g_3(x)\vert (x^ig_2^*(x)-g_2(x))$. \\
 For $c=2+u$, then by (6) we have,
 \begin{equation}
 \begin{split}
 & g_1(x)+(1+u)x^ig_2^*(x)=2+ug_1(x)+(2+u)(1+u)g_2(x)+(1+u)g_3(x)q(x) \\ \nonumber
 & \Rightarrow 3(1+u)g_1(x)+(1+u)x^ig_2^*(x)=3(1+u)g_2(x)+(1+u)g_3(x)q(x) \\
 & \Rightarrow g_1(x)+x^ig_2^*(x)-g_2(x)=g_3(x)q(x). 
 \end{split}
 \end{equation}
 Which implies $g_1(x)+x^ig_2^*(x)-g_2(x) \in (g_3(x))$. Hence, $g_3(x)\vert (g_1(x)+x^ig_2^*(x)-g_2(x))$. Therefore, $g_3(x)\vert (x^ig_2^*(x)-g_2(x))$ as $g_3(x) \vert g_1(x)$, where $i=~deg~g_1(x)-deg~g_2(x)$. 
 Similarly for $c=1+2u$ or $c=2+3u$ we get the above result. \\
 Conversely, for $C$ to be reversible it is sufficient to show that both $(g_1(x)+(1+u)g_2(x))^*$ and $((1+u)g_3(x))^*$ are in $C$. Since $g_3(x)$ is self-reciprocal then $(1+u)g_3(x)\in C$. Also
 \begin{align}
 (g_1(x)+(1+u)g_2(x))^* & = g_1^*(x)+(1+u)x^ig_2^*(x) ~~~~(i=~deg~f_1(x)-deg~f_2(x)) \nonumber \\ 
 & = (g_1(x)+(1+u)g_2(x))+(1+u)x^ig_2^*(x)-(1+u)g_2(x) \nonumber \\ 
 &=(g_1(x)+(1+u)g_2(x))+(1+u)g_3(x)b(x) \in C. \nonumber
 \end{align}
 Therefore, $C$ is reversible.

\vskip 30pt
\centerline {{\hoato 4. Cyclic Reversible Complement Codes over $R$}}

\vskip 6pt
In this section, cyclic codes of odd length over $R$ satisfy the reverse-complement are examined.
First, we give some useful lemmas which can be easily check.

\vskip 6pt
\noindent {\bf Lemma 4.1.}
	{\it For any $a \in R$ we have $a+\overline{a}=1+u$.
}

\vskip 6pt
\noindent {\bf Lemma 4.2.}
{\it For any $a,b,c \in R$, then
	\begin{enumerate}[{\rm (i)}]
		\item $\overline{a+b}=\overline{a}+\overline{b}+3(1+u)$;
		\item $\overline{a+(1+u)b}=\overline{a}+3(1+u)b$.
	\end{enumerate}
}

\vskip 6pt
\noindent {\bf Lemma 4.3.}
{\it For any $a \in R$, we have $\overline{a}+3(1+u)=3a$.
}

\vskip 6pt
\noindent {\bf Theorem 4.4.}
{\it Let $C=\langle g_1(x)+(1+u)g_2(x) \rangle$ be a cyclic code of odd length $n$ over $R$. Then $C$ is a reverse-complement if and only if
	\begin{enumerate}[{\rm (i)}]
		\item $g_1(x)$ is self-reciprocal and $3(1+u)((1-x^n)/(1-x)) \in C$;
		\item \begin{enumerate}[{\rm (a)}]
			\item $x^ig_2^*(x)=g_2(x)$. Or,
			\item $g_1(x)=x^ig_2^*(x)+g_2(x),
			~\text{where}~ i=~deg~g_1(x)-deg~g_2(x)$.
		\end{enumerate}
	\end{enumerate}	
}

\vskip 6pt
\pf Let $C=\langle g_1(x)+(1+u)g_2(x) \rangle$ be the cyclic code of odd length $n$ over $R$. Since the zero codeword must be in $C$, by hypothesis, the WCC of it should also be in $C$. But by Lemma 4.1., we have
\begin{equation}
3~\overline{(0,0,\ldots,0)}=3(1+u,1+u,\ldots,1+u)=3(1+u)\frac{1-x^n}{1-x}\in C.\nonumber
\end{equation}
Now, let
\begin{align}
&
g_1(x)=1+a_1x+\cdots+a_{s-1}x^{s-1}+x^s, \nonumber \\
& ~\text{and}~~~
g_2(x)=1+b_1x+\cdots+b_{r-1}x^{r-1}+x^r, \nonumber 
\end{align}
where $s>r$. Then
\begin{multline}
\lefteqn {g_1(x)+(1+u)g_2(x)} \\ =  (1+a_1x+\cdots+a_{s-1}x^{s-1}+x^s) +  (1+u)(1+b_1x+\cdots+b_{r-1}x^{r-1}+x^r) \\  = (2+u)+(a_1+(1+u)b_1)x + \cdots +  (a_{r-1}+(1+u)b_{r-1})x^{r-1}+ (a_r+(1+u))x^r \\ + a_{r+1}x^{r+1} + \cdots + a_{s-1}x^{s-1} + x^s. \nonumber
\end{multline}
Hence,
\begin{align}
\lefteqn (~{g_1(x)+(1+u)g_2(x)})^{rc}  = & (1+u)(1+x+\cdots+x^{n-s-2}) + ux^{n-s-1} + \overline{a_{s-1}}x^{n-s} + \cdots \nonumber \\ & +\overline{a_{r+1}}x^{n-r-2}  + \overline{(a_r+(1+u))}x^{n-r-1} + \overline{(a_{r-1}+(1+u)b_{r-1})}x^{n-r}\nonumber \\ & + \cdots +\overline{(a_1+(1+u)b_1)}x^{n-2} + \overline{(2+u)}x^{n-1} \nonumber \\ 
= &(1+u)(1+x+\cdots+x^{n-s-2}) + ux^{n-s-1} + \overline{a_{s-1}}x^{n-s} + \cdots \nonumber \\ & +\overline{a_{r+1}}x^{n-r-2} + \overline{a_r}x^{n-r-1}+3(1+u)x^{n-r-1}+\overline{a_{r-1}}x^{n-r} \nonumber \\ & +3(1+u)b_{r-1}x^{n-r}+\cdots+ \overline{a_1}x^{n-2}+(1+u)b_1x^{n-2}+3x^{n-1} \nonumber\\
= & (1+u)(1+x+\cdots+x^{n-s-2}) + ux^{n-s-1} + \overline{a_{s-1}}x^{n-s} + \cdots \nonumber \\ & +\overline{a_{r+1}}x^{n-r-2}+\overline{a_r}x^{n-r-1}+\cdots+ \overline{a_1}x^{n-2} +ux^{n-1}+3(1+u)x^{n-r-1} \nonumber \\ & +3(1+u)b_{r-1}x^{n-r}+\cdots+(1+u)b_1x^{n-2}+3(1+u)x^{n-1} \in C. \nonumber
\end{align}
Since $C$ is linear code, we must have
\begin{equation}
({g_1(x)+(1+u)g_2(x)})^{rc} + 3(1+u)\frac{1-x^n}{1-x}\in C.
\end{equation}
Then, from (7) we have
\begin{multline}
({g_1(x)+(1+u)g_2(x)})^{rc} + 3(1+u)((1-x^n)/(1-x)) \\ = 
3x^{n-s-1} + (\overline{a_{s-1}}+3(1+u))x^{n-s} + \cdots + (\overline{a_1}x^{n-2}+3(1+u))x^{n-2} + 3x^{n-1} \\ +3(1+u)x^{n-r-1}  +3(1+u)b_{r-1}x^{n-r}+\cdots+(1+u)b_1x^{n-2}+3(1+u)x^{n-1} \\ =
3x^{n-s-1}(1+a_{s-1}x+\cdots+a_1x^{s-1}+x^s)\\
3(1+u)x^{n-r-1}(1+b_{r-1}x+\cdots+b_1x^{r-1}+x^r) \\= 
3x^{n-s-1}g_1^*(x) + 3(1+u)x^{n-r-1}g_2^*(x) \\=
3x^{n-s-1}(g_1^*(x)+(1+u)x^{s-r}g_2^*(x)) \nonumber.
\end{multline}
Whence, 
\begin{equation}
g_1^*(x)+(1+u)x^{s-r}g_2^*(x) \in C. \nonumber
\end{equation}
So we have,
\begin{equation}
g_1^*(x)+(1+u)x^{s-r}g_2^*(x)=(g_1(x)+(1+u)g_2(x))a(x). \nonumber
\end{equation}
It is easy to see that the value of $a(x)$ may be $1,3,u,3u,1+2u,2+u,2+3u ~\text{or}~3+2u$. So by previous Theorem 3.3., $g_1(x)=g_1^*(x)$ i.e., $g_1(x)$ s self-reciprocal. Also we have, $x^ig_2^*(x)=g_2(x)$ and $g_1(x)=x^ig_2^*(x)+g_2(x)$, where $i=s-r$.\\
On the other hand, let $c(x)\in C$, then $c(x)=(g_1(x)+(1+u)g_2(x))a(x)$. Since $g_1(x)$ is self-reciprocal and also $x^ig_2^*(x)=g_2(x)$ and $g_1(x)=x^ig_2^*(x)+g_2(x)$, where $i=~deg~g_1(x)-deg~g_2(x)$, then we can write, 
\begin{align}
c^*(x) & =((g_1(x)+(1+u)g_2(x))a(x))^* \nonumber \\ &=(g_1^*(x)+(1+u)x^ig_2^*(x))a^*(x) \nonumber \\ 
& = (g_1(x)+(1+u)x^ig_2^*(x))a^*(x) \nonumber \\ 
& = (g_1(x)+(1+u)x^ig_2^*(x)).c.a^*(x), \nonumber
\end{align}
where $c=1,3,u,3u,1+2u,2+u,2+3u ~\text{or}~3+2u$. Therefore, $c^*(x) \in C$. \\ 
Since $3(1+u)((1-x^n)/(1-x)) \in C$, we have
\begin{equation}
3(1+u)\frac{1-x^n}{1-x}=3(1+u)+3(1+u)x+\cdots+3(1+u)x^{n-1} \in C. \nonumber
\end{equation}
Let $c(x)=c_0+c_1x+\cdots+c_sx^s \in C$. As $C$ is a cyclic code of length $n$, we have
\begin{equation}
x^{n-s-1}c(x)=c_0x^{n-s-1}+c_1x^{n-s}+\cdots+c_sx^{n-1} \in C. \nonumber
\end{equation}
Whence, 
\begin{multline}
\lefteqn ~ 3(1+u)+3(1+u)x+\cdots+3(1+u)x^{n-s-2}+(c_0+3(1+u))x^{n-s-1} \\ +(c_1+3(1+u))x^{n-s}+\cdots+(c_s+3(1+u))x^{n-1} \\ =
3(1+u)+3(1+u)x+\cdots+3(1+u)x^{n-s-2}+ 3\overline{c_0}x^{n-s-1} \\ +3\overline{c_1}x^{n-s}+\cdots+ 3\overline{c_s}x^{n-1}  \\ =
3((1+u)+(1+u)x+\cdots+(1+u)x^{n-s-2}+ \overline{c_0}x^{n-s-1} \\ +\overline{c_1}x^{n-s}+\cdots+ \overline{c_s}x^{n-1}) \in C. \nonumber
\end{multline}
We have $c^*(x)^{rc} \in C$, therefore, $(c^*(x)^{rc})^*=c(x)^{rc} \in C$.

\vskip 6pt
\noindent {\bf Theorem 4.5.}
{\it Let $C=\langle g_1(x)+(1+u)g_2(x),(1+u)g_3(x) \rangle$
	with $g_3(x)\vert g_1(x)\vert \xn$ be a cyclic code of odd length $n$ over $R$. Then $C$ is reverse-complement if and only if
	\begin{enumerate}[{\rm (i)}]
		\item $3(1+u)((1-x^n)/(1-x)) \in C$ and $g_1(x)$ and $g_3(x)$ are self-reciprocal;
		\item $g_3(x)\vert (x^ig_2^*(x)-g_2(x))$, where $i=~deg~g_1(x)-deg~g_2(x).$
	\end{enumerate}
}

\vskip 6pt
\pf Suppose Let $C=\langle g_1(x)+(1+u)g_2(x),(1+u)g_3(x) \rangle$ with $g_3(x)\vert g_1(x)\vert \xn$ be a cyclic code of odd length $n$ over $R$. Since the zero codeword must be in $C$, by hypothesis, the WCC of it should also be in $C$, i.e.,
\begin{equation}
3~\overline{(0,0,\ldots,0)}=3(1+u,1+u,\ldots,1+u)=3(1+u)\frac{1-x^n}{1-x}\in C.\nonumber
\end{equation}
Proceeding in the same way as previous theorem, we get
\begin{equation}
g_1^*(x)+(1+u)x^{s-r}g_2^*(x) \in C. \nonumber
\end{equation}
Where $deg~g_1(x)=s, deg~g_3(x)=r$. 
So we write,
\begin{align}
g_1^*(x)+(1+u)x^{s-r}g_2^*(x) & =(g_1(x)+(1+u)g_2(x))p(x)+ (1+u)g_3(x)q(x).
\end{align}
Here, we get the value of $p(x)$ are $1,3,u,3u,1+2u,2+u,2+3u ~\text{or}~3+2u$. So by Theorem 3.4., $g_1(x)=g_1^*(x)$ i.e., $g_1(x)$ is self-reciprocal. Also we have, $g_3(x)\vert (x^ig_2^*(x)- g_2(x))$, where $i=s-r$.
Now suppose, 
\begin{align}
(1+u)g_3(x)=(1+u)+(1+u)d_1x+\cdots+(1+u)d_{k-1}x^{k-1}+(1+u)x^k. \nonumber 
\end{align}
Then
\begin{multline}
\lefteqn ~(1+u)g_3(x)^{rc} \\ = 
(1+u)(1+x+\cdots+x^{n-k-2}) + x^{n-k-1} + \overline{(1+u)d_{k-1}}x^{n-k} + \cdots + \overline{(1+u)d_1}x^{n-2} \in C \nonumber
\end{multline}
Since $C$ is linear code and $3(1+u)((1-x^n)/(1-x)) \in C$, we must have
\begin{equation}
(1+u)g_3(x)^{rc} + 3(1+u)\frac{1-x^n}{1-x}\in C.\nonumber
\end{equation}
Hence,
\begin{multline}
(1+u)g_3(x)^{rc} + 3(1+u)((1-x^n)/(1-x)) \\ = 
3(1+u)x^{n-k-1} + (\overline{(1+u)d_{k-1}}+3(1+u))x^{n-k} + \cdots + (\overline{(1+u)d_1}+3(1+u))x^{n-2} + 3(1+u)x^{n-1}  \\ =
3ux^{n-k-1} + 3(1+u)d_{k-1}x^{n-k} + \cdots + 3(1+u)d_1x^{n-2} + 3(1+u)x^{n-1}  \\ =
3x^{n-k-1}((1+u)+(1+u)d_{k-1}x+\cdots+(1+u)d_1x^{k-1}+(1+u)x^k) \\ =
3x^{n-k-1}(1+u)g_3^*(x) \in C \nonumber.
\end{multline}
Therefore, $g_3(x)=g_3^*(x)$.\\
Conversely, let $c(x)\in C$, then $c(x)=(g_1(x)+(1+u)g_2(x))p(x)+(1+u)g_3(x)q(x)$. Since $g_1(x)$ and $g_3(x)$ are self-reciprocal, $g_3(x)\vert (x^ig_2^*(x)-g_2(x))$, 
where $i=~deg~g_1(x)-deg~g_2(x)$,\\ then we write, 
\begin{align}
c^*(x) & =((g_1(x)+(1+u)g_2(x))p(x))^*+((1+u)g_3(x)q(x))^* \nonumber \\ &=(g_1^*(x)+(1+u)x^ig_2^*(x))p^*(x)+(1+u)x^jg_3^*(x)q^*(x) \nonumber \\ 
& = (g_1(x)+(1+u)g_2(x))p^*(x)+ (1+u)m^*(x)(x^ig_2^*(x)-g_2(x)) + (1+u)x^jg_3(x)q^*(x)  \nonumber \\ 
& = (g_1(x)+(1+u)g_2(x))p^*(x) + (1+u)g_3(x)m(x) \nonumber
\end{align}
Where $i=~deg~g_1(x)-deg~g_2(x)$ and $j=~deg~(g_1(x)p(x))-deg~(g_3(x)q(x))$. \\Therefore, $c^*(x) \in C$. \\ 
As $3(1+u)((1-x^n)/(1-x)) \in C$, we have
\begin{equation}
3(1+u)\frac{1-x^n}{1-x}=3(1+u)+3(1+u)x+\cdots+3(1+u)x^{n-1} \in C. \nonumber
\end{equation}
Let $c(x)=c_0+c_1x+\cdots+c_sx^s \in C$. As $C$ is a cyclic code of length $n$, we have
\begin{equation}
x^{n-s-1}c(x)=c_0x^{n-s-1}+c_1x^{n-s}+\cdots+c_sx^{n-1} \in C. \nonumber
\end{equation}
Hence, 
\begin{multline}
\lefteqn ~ 3(1+u)+3(1+u)x+\cdots+3(1+u)x^{n-s-2}+(c_0+3(1+u))x^{n-s-1} \\ +(c_1+3(1+u))x^{n-s}+\cdots+(c_s+3(1+u))x^{n-1} \\ =
3(1+u)+3(1+u)x+\cdots+3(1+u)x^{n-s-2}+ 3\overline{c_0}x^{n-s-1} \\ +3\overline{c_1}x^{n-s}+\cdots+ 3\overline{c_s}x^{n-1}  \\ =
3((1+u)+(1+u)x+\cdots+(1+u)x^{n-s-2}+ \overline{c_0}x^{n-s-1} \\ +\overline{c_1}x^{n-s}+\cdots+ \overline{c_s}x^{n-1}) \in C. \nonumber
\end{multline}
Therefore, $c^*(x)^{rc} \in C$ and $(c^*(x)^{rc})^*=c(x)^{rc} \in C$.

\vskip 30pt
\centerline {{\hoato 5. The $GC$-content and the deletion distance $D$}}

\vskip 6pt
In molecular biology, $GC$-content (or guanine-cytosine content) is the percentage of nitrogenous bases on a DNA molecule that are either guanine($G$) or cytosine($C$). The $G$ and $C$ pair need three hydrogen bonds, while $A$ and $T$ pairs need two hydrogen bonds. Since the chemical bonds between the WCC pairs are different and the total energy of the DNA molecule depends on the number of $A$ and $T$ pairs and the number of $C$ and $G$ pairs. As DNA with high energy $GC$-content is more stable than DNA with low energy $GC$-content, it is always desirable in a DNA code to have all codewords with the same $GC$-content, so that they have similar melting temperatures.
In this section we study the $GC$-content of a DNA cyclic code and its deletion distance $D$.

\vskip 6pt
For two quaternary $n$-sequences $X$ and $Y$ the energy of DNA hybridization $E(X,Y)$ is the longest common subsequence of either strand or the reverse complement of the other strand. We define the \emph{deletion similarity} $S(X,Y)$ as the length of the longest common subsequence (LCS) for $X$ and $Y$. Note that for any strands $X$ and $Y$ of length $n$ we have 
$$ S(X,X)=n ~~\text{and}~~S(X,Y)=S(Y,X).$$
The number of base pair bonds between $X$ and $Y^{rc}$ recognize the deletion similarity $S(X,Y)$, i.e., we have
$$ E(X,Y^{rc})=E(Y^{rc},X)=S(X,Y)=S(Y,X).$$

\vskip 6pt
\noindent {\bf Example 5.1.}
	{\it If $X=TCAGG$ and $Y=TACGT$ then deletion similarity $S(X,Y)=3$ as $TCG$ is the longest common subsequence (LCS) for both $X$ and $Y$. Note that $TCG$ is not unique since $TAG$ is another common subsequence of length $3$.
}
\vskip 6pt
\begin{definition}
	The \emph{Hamming weight enumerator}, $W_C(y)$, of a code $C$ is defined as 
	$$ W_C(y)=\sum_i A_iy^i,$$
	Where $A_i=|~ \{c \in C~|~w(c)=i\} ~|$, i.e., the number of codewords in $C$ whose weights equal to i. The smallest non-zero exponent of $y$ with a non-zero coefficient in $W_C(y)$ is equal to the
	minimum Hamming distance of the code.
\end{definition}

\begin{definition}\textnormal{\cite{si}}
	For any cyclic code $C=(g_1(x)+(1+u)g_2(x),(1+u)g_3(x))$, define the subcode $C_{1+u}$ to consist of all codewords in $C$ that are multiples of $(1+u)$.
\end{definition}

\noindent {\bf Theorem 5.2.}
	{\it Let $C=(g_1(x)+(1+u)g_2(x),(1+u)g_3(x))$ be a cyclic code of odd length $n$. Then, $C_{1+u}=((1+u)g_3(x))$.
}

\vskip 6pt
\pf We note that \\
$~~~~~~~~~~~~~~~~~~((1+u)g_3(x))\subseteq C_{1+u}$.\\
Now we have to show that $C_{1+u} \subseteq ((1+u)g_3(x))$. Let $c \in C$. Then,
$$ c(x)=(g_1(x)+(1+u)g_2(x))r(x)+(1+u)g_1(x)s(x)+(1+u)g_3(x)t(x),$$
where $r(x),s(x),t(x) \in \Z_4[x]$. If $c$ is multiple of $(1+u)$, then we must have\\ $\xn \vert (g_1(x)+(1+u)g_2(x))r(x)$ and hence
$$ c(x)=(1+u)g_1(x)s(x)+(1+u)g_3(x)t(x).$$
Since $g_3(x) \vert g_1(x)$, then $c(x) \in (1+u)g_3(x)$ and hence $C_{1+u} \subseteq ((1+u)g_3(x))$.\\
Therefore, $C_{1+u}=((1+u)g_3(x))$. \\
We get the following theorem from \cite{si}.

\vskip 6pt
\noindent {\bf Theorem 5.3.}
	{\it Let $C=(g_1(x)+(1+u)g_2(x),(1+u)g_3(x))$ be a cyclic code of odd length $n$. Let $c \in C$, with $\omega(c)=w_H(\overline{c})$ where $\overline{c} \in Z_4[x]$. Further, all possible spectra of the $GC$-content of $C$ are
	determined by the Hamming weight enumerator of the binary code generated by $a(x)$.
}

\vskip 6pt
\noindent {\bf Theorem 5.4.}
	{\it Let $C=(g_1(x)+(1+u)g_2(x),(1+u)g_3(x))$ be an $(n,D)$ cyclic code of odd length $n$ and $D_{1+u}$ be the deletion similarity distance of the code $C_{1+u}$. Then $D=D_{1+u}$.
}

\vskip 6pt
\pf Since $C_{1+u}\subseteq C$, then for any $X,Y \in C$ we get \\
$~~~~~~~~~~~~~~~~~~~~~~~~~~~~S(X,Y)\leq n-D-1.$ \\
Which implies $D_{1+u}\geq D$. Now suppose there are two codewords $A, B \in C$ such that \\ 
$~~~~~~~~~~~~~~~~~~~~~~~~~~~~S(A,B)> n-D_{1+u}-1.$\\
From Theorem 5.2., as $A$ and $B$ are in $C$, then $(1+u)A$ and $(1+u)B$ are two codewords in $C_{1+u}$. So it is clear that\\ 
$~~~~~~~~~~~~~~~~~~~~~~~~~~~~S((1+u)A,(1+u)B)>S(A,B)> n-D_{1+u}-1,$\\
which contradict. Therefore, $D=D_{1+u}$.

\vskip 30pt
\centerline {{\hoato 6. Example}}

\vskip 6pt
In this section, we give some examples of cyclic codes of different lengths over the ring $R$ to illustrate the above results. 

\vskip 6pt
\noindent {\bf Example 6.1.}
	{\it Let
	$$ x^3-1 = (x-1)(x^2 + x +1)=f_1f_2 \in \Z_4[x].$$
	\begin{enumerate}
		\item Let $C=\langle g_1(x)+(1+u)g_2(x) \rangle$, where $g_1(x)=g_2(x)=f_2=(x^2 + x +1)$. It is easy to check that $g_1(x)$ is self-reciprocal and $x^ig_2^*(x)=g_2(x)$, where $i=~deg~g_1(x)-deg~g_2(x)$. $C$ is a cyclic DNA
		code of length 3 with R-C property and minimum Hamming distance 3. The image of $C$ under the map $\Phi$ is a DNA code of length 6, size 16 and minimum Hamming distance 3. These codewords are given in Table 2.\\
		\item Let $C=\langle g_1(x)+(1+u)g_2(x),(1+u)g_3(x) \rangle$, where $g_1(x)=g_2(x)=f_1f_2$ and $g_3(x)=f_2$. We check that $g_1(x)$ and $g_3(x)$ are self-reciprocal and $g_3(x)\vert ((1+u)x^ig_2^*(x)+(1+u)g_2(x))$. Then $C$ is a cyclic DNA code of length 3 with R-C property and minimum Hamming distance 3. The image of $C$ under the map $\Phi$ is a DNA code of length 6, size 64 and minimum Hamming distance 2.
	\end{enumerate} 
}
\begin{center}
	{\bf Table 2.} A DNA code of length 6 obtained from \\ $C=\langle (x^2 + x +1)+2(x^2 + x +1) \rangle$\\~\\
	\begin{tabular}{ l  c  c  c }
		\hline
		
		$AAAAAA$ &  $TTTTTT$ &  $CCCCCC$ &  $GGGGGG$  \\
		$ATATAT$ &  $TATATA$ &  $CTCTCT$ &  $GAGAGA$  \\
		$AGAGAG$ &  $TCTCTC$ &  $CGCGCG$ &  $GCGCGC$   \\
		$ACACAC$ &  $TGTGTG$ &  $CACACA$ &  $GTGTGT$   \\
		\hline
		
	\end{tabular}
\end{center}
Here, $X\neq X^{rc}$ for all $X \in C$. Further, $S(X,Y) \in \{ 0,3 \}~\forall~X,Y \in C$ and $X \neq Y$. Since the image of $C$ under the map $\Phi$ is a DNA code of length 6 and $S(X,Y)\leq 3$, we get $D=2$. So, this is an $(6,2)$ DNA cyclic code.
\vskip 6pt
\noindent {\bf Example 6.2.}
	{\it Let
	$$ x^9-1 = (x-1)(x^2 + x +1)(x^6 + x^3 + 1)=f_1f_2f_3 \in \Z_4[x].$$
	Let $C=\langle g_1(x)+(1+u)g_2(x) \rangle$, where $g_1(x)=g_2(x)=f_2f_3$. It is easy to check that $g_1(x)$ is self-reciprocal and $x^ig_2^*(x)=g_2(x)$, where $i=~deg~g_1(x)-deg~g_2(x)$. $C$ is a cyclic DNA
	code of length 9 with R-C property and minimum Hamming distance 9. The image of $C$ under the map $\Phi$ is a DNA code of length 18 and minimum Hamming distance 9. The number of codewords are 16 which are given in Table 3.
}

\begin{center}
	{\bf Table 3.} A DNA code of length 18 obtained from the above code.\\~\\
	\begin{tabular}{ l  c   }
		\hline
		
		$AAAAAAAAAAAAAAAAAA$ &  $TTTTTTTTTTTTTTTTTT$ \\ $CCCCCCCCCCCCCCCCCC$ &  $GGGGGGGGGGGGGGGGGG$  \\
		$ATATATATATATATATAT$ &  $TATATATATATATATATA$ \\  $CTCTCTCTCTCTCTCTCT$ &  $GAGAGAGAGAGAGAGAGA$  \\
		$AGAGAGAGAGAGAGAGAG$ &  $TCTCTCTCTCTCTCTCTC$ \\ $CGCGCGCGCGCGCGCGCG$ &  $GCGCGCGCGCGCGCGCGC$   \\
		$ACACACACACACACACAC$ &  $TGTGTGTGTGTGTGTGTG$ \\  $CACACACACACACACACA$ &  $GTGTGTGTGTGTGTGTGT$   \\
		\hline
		
	\end{tabular}
\end{center}	
Here, $X\neq X^{rc}$ for all $X \in C$. Further, $S(X,Y) \in \{ 0,9 \}~\text{for all}~X,Y \in C$ and $X \neq Y$. Since the image of $C$ under the map $\Phi$ is a DNA code of length 18 and $S(X,Y)\leq 3$, we get $D=8$. So, this is an $(18,8)$ DNA cyclic code.

\vskip 30pt
\centerline {{\hoato 7. Conclusion}}

\vskip 6pt
This paper constructed codes over an alphabet $\{A, C, G, T \}$ relevant to the design of synthetic DNA
strands used in DNA microarrays, as DNA tags in chemical libraries and in DNA computing.
Here, we developed the structure of DNA cyclic codes over the ring $R=\Z_4[u]/\langle u^2-1 \rangle$ of odd length based on deletion distance which is motivated by the simple and less complexity structure of these codes over $R$. Reversible codes and reverse-complement codes related to cyclic codes are studied, respectively. $GC$-content of these codes and their deletion distance are discussed and proved that the $GC$-content and deletion distance of a code $C$ are related to a subcode of $C$. We add several examples of cyclic DNA codes of different lengths including their $GC$-content and the number of deletion errors a code can correct. For future study, the algebraic structure of cyclic codes of even length and their relation to DNA codes is still an open problem.


\vskip 6pt
\begin {thebibliography}{100}

\bibitem{ab} Abualrub T., Ghrayeb A., Zeng X., Construction of cyclic codes over $GF(4)$ for DNA computing, \emph{J. Franklin Institute}, {\bf 343}(4-5), 448-457, (2006).

\bibitem{ab2} Abualrub T., Siap I., Cyclic codes over the rings $\Z_2 + u\Z_2$ and $\Z_2 + u\Z_2 + u^2\Z_2$, \emph{Des. Codes Cryptogr.}, {\bf 42}, 273-287, (2007).

\bibitem{ad} Adleman L., Molecular computation of the solution to combinatorial problems, \emph{Science}, {\bf 266}, 1021-1024, (1994).

\bibitem{ba} Bayram A., Oztas E. S., Siap I., Codes over $\F_v+u\F_4$ and some DNA applications, \emph{Designs,Codes and Cryptography}, DOI 10.1007/s10623-015-0100-8, (2015).

\bibitem{be} Bennenni N., Guenda K., Mesnager S., New DNA cyclic codes over rings, arXiv: 1505.06263v1 [cs.IT], (2015).


\bibitem{Dy}  D'yachkov A., Macula A., Renz T., Vilenkin P., Ismagilov I., New results on DNA codes, \emph{International
Symposiumon Information Theory}, 283-287, (September 2005).

\bibitem{Dy1}  D'yachkov A., Erdos P., Macula A., Torney D., Tung C. H., Vilenkin P., White P. S., Exordium for DNA
codes, \emph{Journal of Combinatorial Optimization}, {\bf 7}, 369-379, (2003).

\bibitem{ga} Gaborit P., King O. D., Linear construction for DNA codes, \emph{Theor. Computer Science}, {\bf 334}(1-3), 99-113, (2005).

\bibitem{gu} Guenda K., Gulliver T. A., Sol\'{e} P., On cyclic DNA codes, \emph{Proc. IEEE Int. Symp. Inform. Theory}, Istanbul, 121-125, (2013). 

\bibitem{gu1} Guenda K., Aaron, Gulliver T., Construction of cyclic codes over $\F_2 + u\F_2$ for DNA
computing, \emph{AAECC}. {\bf 24}, 445-459, (2013).

\bibitem{lia} Liang J., Wang L., On cyclic DNA codes over $\F_2+u\F_2$, \emph{J. Appl. Math. Comput.}, DOI 10.1007/s12190-015-0892-8, (2015).

\bibitem{lip} Lipton R. J., DNA solution of hard computational problems, \emph{Science}, {\bf 268}, 542-545, (1995). 

\bibitem{mar} Marathe A., Condon A. E., Corn R.M., On combinatorial DNA word design, {\emph J. Comput. Biol}, {\bf 8}, 201-220, (2001). 

\bibitem{mas}  Massey J. L., Reversible codes, {\emph Inf. Control}, {\bf 7}, 369-380, (1964).

\bibitem{oze}  Ozena M., Uzekmeka F. Z., Aydin N., Özzaima N. T., Cyclic and some constacyclic codes over the ring $\Z_4[u]/\langle u^2-1 \rangle$, \emph{Finite Fields Appl.}, {\bf 38}, (2016), 27-39. 

\bibitem{ozt} Oztas E. S., Siap I., Lifted polynomials over $\F_{16}$ and their applications to DNA codes, \emph{Filomat}, {\bf 27}, 459-466, (2013).

\bibitem{pa} Pattanayak S., Singh A. K., On cyclic DNA codes over the Ring $\Z_4 + u\Z_4$, arXiv:1508.02015[cs.IT], (2015).

\bibitem{si} Siap I., Abualrub T., Ghrayeb A., Cyclic DNA codes over the ring $\F_2[u]/(u^2-1)$ based on the deletion distance, \emph{J. Franklin Institute}, {\bf 346}, 731-740, (2009).

\bibitem{yi} Yildiz B., Aydin N., On cyclic codes over $\Z_4+u\Z_4$ and their $\Z_4$–images, \emph{Int. J.  Information and Coding Theory}, {\bf 2(4)}, (2014), 226-237.

\bibitem{yi1}  Yildiz B., Karadeniz S., Linear codes over $\Z_4+u\Z_4$, MacWilliams identities, projections, and formally self-dual codes, \emph{Finite Fields Appl.}, {\bf 27}, (2014), 24-40.

\bibitem{yi2} Yildiz B., Siap I., Cyclic DNA codes over the ring $\F_2[u]/(u^4-1)$ and applications to DNA codes, \emph{Comput. Math. Appl}, {\bf 63}, 1169-1176, (2012). 

\bibitem{zh} Zhu S., Chen X., Cyclic DNA codes over $\F_2 +u\F_2 +v\F_2 + uv\F_2$, arXiv: 1508.07113v1 [cs.IT], (2015).

\end {thebibliography}
\end{document}